# Predicting the thermodynamic stability of perovskite oxides using machine learning models

Wei Li [a], Ryan Jacobs [a], Dane Morgan [a] [*]

[a] Department of Materials Science and Engineering, University of Wisconsin-Madison, Madison, Wisconsin, 53706, USA.

[*] Corresponding author, Dane Morgan, email address: ddmorgan@wisc.edu

**Abstract**

Perovskite materials have become ubiquitous in many technologically relevant applications, ranging from catalysts in solid oxide fuel cells to light absorbing layers in solar photovoltaics. The thermodynamic phase stability is a key parameter that broadly governs whether the material is expected to be synthesizable, and whether it may degrade under certain operating conditions. Phase stability can be calculated using Density Functional Theory (DFT), but the significant computational cost makes such calculation potentially prohibitive when screening large numbers of possible compounds. In this work, we developed machine learning models to predict the thermodynamic phase stability of perovskite oxides using a dataset of more than 1900 DFT-calculated perovskite oxide energies. The phase stability was determined using convex hull analysis, with the energy above the convex hull ($E_{hull}$) providing a direct measure of the stability. We generated a set of 791 features based on elemental property data to correlate with the $E_{hull}$ value of each perovskite compound, and found through feature selection that the top 70 features were sufficient to produce the most accurate models without significant overfitting. For classification, the extra trees algorithm achieved the best prediction accuracy of 0.93 (+/- 0.02), with an $F_1$ score of 0.88 (+/- 0.03). For regression, leave-out 20% cross-validation tests with kernel ridge regression achieved the minimal root mean square error (RMSE) of 28.5 (+/- 7.5) meV/atom between cross-validation predicted $E_{hull}$ values and DFT calculations, with the mean absolute error (MAE) in cross-validation energies of 16.7 (+/- 2.3) meV/atom. This error is within the range of errors in DFT formation energies relative to elemental reference states when compared to experiments and therefore may be considered sufficiently accurate to use in place of full DFT calculations. We further validated our model by predicting the stability of compounds not present in the training set and demonstrated our machine learning models are a fast and effective means of obtaining qualitatively useful guidance for a wide-range of perovskite oxide stability, potentially impacting materials design choices in a variety of technological applications.

**Highlights**

- Performed machine-learning based studies on a dataset of DFT-calculated stability data of over 1900 perovskite oxides.
- Demonstrated a complete workflow from feature generation and selection to model validation and testing.
- Showed that a machine learning approach is capable of accurately and efficiently obtaining stability information for a wide composition range of perovskite oxides.
- Showed that a machine learning prediction of perovskite oxide stability can supplement DFT calculations for faster screening of novel materials.

**Main**

**1 Introduction**

The discovery of novel functional materials is central to the continuing development of materials technologies. Recently, high-throughput DFT methods have been used to guide the discovery of new compounds for numerous applications, including: perovskite oxides for solid oxide fuel cell (SOFC) cathodes[1, 2], thermochemical water splitting,[3] half-heusler and sintered compounds for thermoelectrics,[4, 5] oxides and oxynitrides for light harvesting[6] and photoelectrochemical water-splitting[7, 8], and binary metal alloys for electrocatalytic hydrogen evolution[9] and oxygen reduction.[10] While high-throughput DFT studies are valuable for discovering new functional materials, they suffer from the high computational cost required to conduct hundreds to thousands of DFT calculations.

In an effort to reduce the large amount of time required to conduct large-scale screening studies, either computational or experimental, we here apply machine learning approaches that have been demonstrated to efficiently predict many properties of materials given only relatively easily obtained structural or compositional information. Examples of properties predicted using machine learning approaches include: relative permittivity and oxygen diffusion properties of ceramic materials,[11] band gap of inorganic materials,[12] formation energy of elpasolite structures,[13] molecular electronic properties in chemical compound space,[14] density of electronic states at the Fermi energy,[15] molecular atomization energies of molecules,[16] Curie temperature of high-temperature piezoelectric perovskites,[17] thermodynamic stability of ternary oxide compounds,[18] and band gap energy of crystalline compounds and metallic glass-forming ability of ternary amorphous alloys.[19] Accurate machine learning model predictions for a material can be orders of magnitude faster than the corresponding DFT simulations or experiments, allowing them to be used to quickly understand trends in materials properties and inform materials discovery.

Of the numerous materials families investigated with high-throughput DFT methods, perovskite materials stand out as a particularly challenging class of materials for computational screening and property evaluation. When one accounts for the large number of different A- and B-site elements, as well as different typical dopant ratios and combinations, the potential number of unique perovskite compositions may be easily greater than $10^7$ materials (assuming 18 possible A-site species, 31 possible B-site species, and possibly mixing up to 3 components on each site with composition restricted to increments of 0.25). This compositional flexibility of the perovskite structure enables an array of complex functional properties, including active catalysis of many reactions, ferroelectricity, piezoelectricity, superconductivity and efficient light-to-energy conversion. This flexibility also creates a significant challenge to predicting the thermodynamic stability, as stoichiometric alloying information needs to be taken into account for the different sublattices of the $ABX_3$ structure (where A and B are one or more cations and X is one or more anions). Recently, Schmidt, et al. reported their work on the stability prediction of ternary perovskite and anti-perovskite compounds, which used a DFT-generated dataset of about 250,000 $ABX_3$ compounds. The A, B, and X species were chosen from a pool of more than 60 elements ($64 \times 63 \times 62 = 249,984$) and a achieved mean absolute error of 121 meV/atom for regression of energy above the convex hull.[20] However, there are a large number of quaternary or quinary perovskite materials with doped elements in the A- and B- sites in an array of technologically relevant applications, so it is important to also explore the use of machine learning approaches on perovskites which have alloying on the A- and B-sites.

Recently, Jacobs, et al. used high-throughput DFT methods to screen the catalytic activity and thermodynamic phase stability of 2145 perovskite oxides for use as SOFC cathodes. [2] In general, the thermodynamic phase stability of a perovskite is a key materials property, the value of which may determine the utility of the perovskite in the given application of interest. The stability typically correlates at least loosely with whether a perovskite is synthesizable, as well as whether it may be expected to degrade (or remain stable) over time under some operational environment, such as a specific working temperature or partial pressure of oxygen.[2, 17] In the work of Jacobs, et al., the stability of perovskite oxides was evaluated by using the phase diagram tools contained within the Pymatgen toolkit. The phase diagram tools in Pymatgen enable one to perform convex hull analysis, where the stability of a particular material composition (e.g. $LaFeO_3$) within a user-provided composition space (e.g. all inorganic crystalline compounds comprising the La-Fe-O system) can be performed. The main parameter governing stability is the energy above the convex hull ($E_{hull}$).[21] The value of $E_{hull}$ is a measure of the decomposition energy of the compound into a linear combination of the stable phases present on the phase diagram. Thermodynamically stable compounds exhibit an $E_{hull}$ of zero (i.e., they are on the convex hull and are stable, equilibrium phases present on the phase diagram, at least at near zero temperature), and more positive values of $E_{hull}$ indicate decreasing stability.[22] Based on the provided example above for $LaFeO_3$, this material is thermodynamically stable and has $E_{hull} = 0$ meV/atom. However, if one were to dope Sr on the A-site and Co on the B-site of $LaFeO_3$ to create $La_{0.375}Sr_{0.625}Co_{0.25}Fe_{0.75}O_3$ (LSCF, a well-studied commercial SOFC cathode material), then the convex hull analysis of this compound in the La-Sr-Fe-Co-O system results in $E_{hull} = 47$ meV/atom, where the energy is relative to the more stable decomposition products of $LaFeO_3$, $Sr_2Co_2O_5$, $Sr_2Fe_2O_5$, and $O_2$. This analysis indicates that LSCF is less stable than $LaFeO_3$, as the $E_{hull}$ value of LSCF is larger. The pool of approximately 2145 perovskite materials calculated by Jacobs, et al. represents a very small fraction of the composition space of possible perovskite oxide compositions. Thus, data-driven methodologies based on machine learning would be beneficial to predict the stability of many additional perovskite oxide compounds.

In this work, we predict the thermodynamic phase stability of perovskite oxides using machine learning models and a subset of the perovskite stability data from Jacobs, et al.[2] of 1929 compounds (these 1929 were the subset of the 2145 compounds available at the time of writing this paper). The model can serve as a screening tool for fast discovery of potential stable compounds, significantly reducing DFT computational time and effort. We have trained several machine learning models for both classification and regression. For classification of determining stable versus unstable compounds, we found that the extra trees classifier (also known as extremely randomized trees),[23] resulted in the best classification model as determined by its calculated precision, recall and $F_1$ score of stable/unstable predictions. For regression of the $E_{hull}$ values, we found the kernel ridge regression model [24] after parameter optimization gave the best regression fitting performance as determined by its calculated $R^2$ score and RMSE of predicted $E_{hull}$ values. Overall, our model can predict the thermodynamic phase stability of perovskite oxide materials with uncertainties that are within typical DFT energy error bars compared to experiments.

**2 Methods**

The construction and validation of our machine learning models to predict perovskite stability involved five steps: (i) Generation of a feature set that can describe the thermodynamic properties of perovskite oxides. (ii) Identification of relevant features that show high correlation with stability through feature selection. (iii) Selection of the best machine learning model from the set of candidate machine learning algorithms. (iv) Examination of the model validity for different perovskite composition spaces, based on the frequency each element occurs in the training dataset. (v) Prediction of thermal stability of new perovskites outside of the dataset and comparison of the predicted values with DFT calculations. In the following sections, we detail each of the above steps needed to construct our machine learning models.

In this work, we have used the python library scikit-learn[25] for all machine learning models, feature selection methods and model evaluations. Scikit-learn is an open source machine learning package distributed under BSD license. A summary of all scikit-learn routines and function calls used in this work is provided in the **Data in Brief (DiB)** [26]. The training dataset of perovskite oxide compositions and DFT-calculated $E_{hull}$ values, as well as the project source code and best models are also provided in the **DiB**.

**2.1 Dataset and feature generation**

The training dataset was comprised of 1929 perovskite oxide compositions from the work of Jacobs, et al.[2] These perovskite materials were simulated using DFT methods, and the stability of each compound was analyzed using the Pymatgen toolkit and all DFT-calculated materials present in the Materials Project online database as of December 2016.[22] The $E_{hull}$ values were obtained under environmental conditions of T = 1073 K, $p(O_2)$ = 0.2 atm (this corresponds to an oxygen chemical potential of -6.25 eV/O, which is –1.31 eV/O relative to the $O_2$ molecule energy calculated in the Materials Project (material identification number mp-12957)), which represents the approximate working conditions of SOFC cathodes. Additionally, $H_2$ was present in the phase stability calculations via equilibrium with $O_2$ and $H_2O$ gas, and a relative humidity of 30%. Additional computational details can be found in Jacobs, et al.[2] We note here that based on our choice of *T* and $p(O_2)$ conditions, the present model is suitable for predicting the stability of perovskites at elevated temperature at approximately room $p(O_2)$ conditions. We believe this choice of thermodynamic conditions does not overly limit the general applicability of our model in predicting perovskite stability because (1) many technological applications involving the use of perovskite oxides operate at elevated temperatures or in environments that are otherwise more reducing that standard conditions, and (2) the examination of stability at 1073 K is a relevant temperature regime where cation motion may be sufficiently fast to result in phase decomposition that is accurately predicted by the thermodynamic methods employed here, that is, the decomposition of these materials is not likely to be kinetically limited. The perovskite compounds consist of elements from a candidate set of A= {Ba, La, Y, Pr, Gd, Dy, Ho, Nd, Sm, Ca, Sr, Bi, Cd, Sn, Zn} and B= {Fe, V, Cr, Mn, Sc, Co, Ti, Mg, Ni, Zr, Ga, Hf, Nb, Ta, Re, Tc, Ir, Os, Ru, Rh, Al, Cu, Pt, Zn}, and X = O. In total, the dataset consists of 71 ternary (e.g. $LaMnO_3$), 1248 quaternary (e.g. $La_{0.75}Sr_{0.25}MnO_3$), 601 quinary (e.g. $La_{0.5}Sr_{0.5}Co_{0.25}Fe_{0.75}O_3$) and 9 sextenary perovskite oxides. The perovskite materials used in this work, which were taken from the work of Jacobs, et al. were all modeled as fully stoichiometric (no O vacancies), except for a few select cases where the O nonstoichiometry was explicitly known from experiment.[2] All compounds containing more than 3 elements (including O) were simulated as a single ordered structure (some ordering was forced by the use of relatively small periodic supercells) so that the large number of DFT calculations was tractable. Given the size of the simulation supercells used in the work of Jacobs, et al., the A- and B-site alloying compositions were binned at 1/8 site fraction, up to 50% alloying on a particular sublattice. Chemical orderings on each sublattice within each supercell made by positioning elements of the same type as far apart as possible. For additional details, the reader is directed to the methods section of the work of Jacobs, et al.[2] We have used an $E_{hull}$ value of 40 meV/atom as the threshold to separate stable and unstable perovskites (i.e., $E_{hull}$ <= 40 meV/atom is a stable compound). We note here that any compound that is not on the convex hull is technically unstable. However, previous analysis of typical DFT errors using compounds in the Materials Project database has suggested that 40 meV/atom is a reasonable cutoff to separate materials that are most likely stable versus those that may be metastable or unstable.[7, 27] In the work of Wu, et al.[7] the DFT error bar value from convex hull analysis was obtained by examining the DFT-calculated $E_{hull}$ values of compounds contained in the ICSD, which have all been experimentally synthesized. Wu, et al. found that more than 80% of the compounds in the ICSD have an $E_{hull}$ value of less than 36 meV/atom, which we have rounded up to the 40 meV/atom benchmark used in this work. **Fig. 1** shows the histogram of $E_{hull}$ values of all perovskites in the dataset, among which 567 compounds are stable perovskites and 1362 compounds are unstable perovskites. We note that the values of $E_{hull}$ and the categorical value of stable/unstable, which will both be the focus of this paper, are specific to a given chemical potential of oxygen. To allow users to work at other oxygen chemical potentials, we have also included a data set and associated machine learning model for formation energies relative to pure elements and oxygen gas at the above conditions. These formation energies and the predictions of the associated machine learning model can be readily transformed to any oxygen chemical potential by a constant shift of 1.31 eV/O, as discussed above. The best regression model trained for formation energy prediction demonstrated a cross-validation $R^2$ of 0.988, RMSE score of 0.062 eV/atom and MAE score of 0.032 eV/atom. The formation energy data and associated machine learning model discussion can be found in **Sec. S3** of the supplemental information in **DiB**.

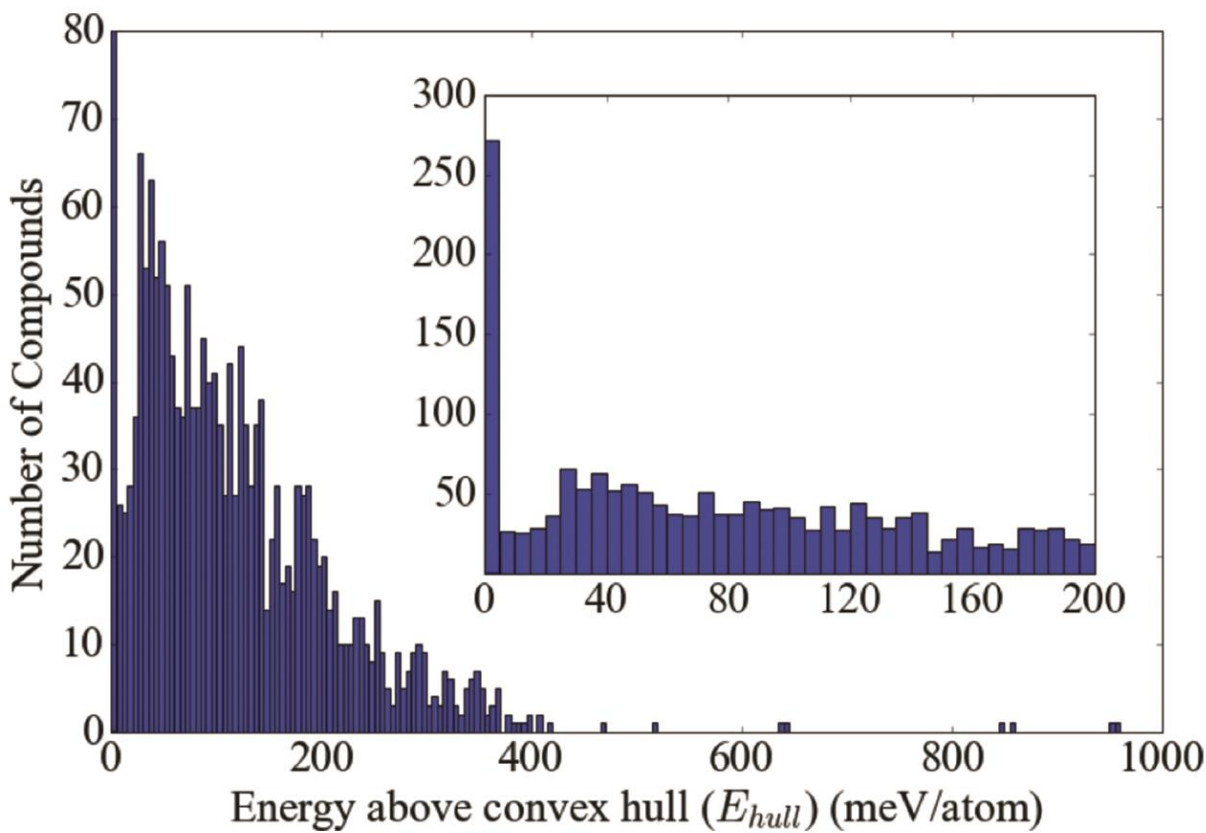

**Fig. 1** $E_{hull}$ distribution of all 1929 perovskite compounds. The bin size in the histogram is 5 meV/atom. The inset plot shows a zoomed-in region from 0 to 200 meV/atom of the main plot. The number of compounds residing on the convex hull is over 250, as shown in the inset plot.

To construct the matrix of features used to train our machine learning models, we used an expansive elemental property database of physical and chemical properties of elements in their atomic form as compiled from the Materials Agnostic Platform for Informatics and Exploration (MAGPIE)[19] database and the web chemical elements database in Resources for Teaching Science.[28] First, we generated features using the elemental properties of the highest composition element on each of the A, B, and X sites (e.g., if we had $La_{0.75}Sr_{0.25}Co_{0.80}Fe_{0.2}O_3$, we made a new feature that was just the elemental properties of La and then of Co). Then, considering that the alloying elements and number of elements in the perovskite A- and B-sites varies, we created additional features by calculating the maximum, minimum, difference, and weighted average by atomic fraction in each site for every physical and chemical property in the elemental property database. This approach assures we have the same number of features that specifically incorporate stoichiometric information for each compound, regardless of the number of elements contained in the composition. In addition, we generated the following new features to describe structural characteristics unique to perovskite materials: Goldschmidt tolerance factor,[29] octahedral factor,[30] and A-O, B-O bond length using composition-averaged Shannon radii.[31] After assembling the complete set of features, there were a total of 962 features, 529 of which were continuous features and 433 of which were discrete features. The complete training dataset and feature matrix is provided in a spreadsheet as part of the **DiB**.

**2.2 Feature selection**

We have tested three feature selection methods in order to remove redundant or irrelevant features: stability selection,[32] recursive feature elimination (RFE),[33] and univariate feature selection based on mutual information.[34] In stability selection, features are selected based on the fraction of times that the randomized procedure picks a given feature by repeating random subsamples of the data and fitting to a logistic regression model in classification task (or lasso model for regression).[35] RFE selects the most relevant features by recursively removing those features which exhibit the smallest weight as assigned by an extra trees classifier in classification task (or extra trees regressor for regression).[23] Univariate feature selection ranks all features by the amount of mutual information between each feature and the target classification value. The mutual information measures the degree of dependence between features and the target value based on entropy estimated by a nearest neighbor method.[34]

All features are normalized to have zero mean and standard deviation of unity prior to performing feature selection. This normalization is used to ensure that all features are scaled in the same way, as standardization of the feature set is a common requirement for many machine learning models, such as artificial neural networks[36] and support vector machines,[37] which are both sensitive to feature scaling. We performed feature selection for both classification and regression tasks. For classification, we used 20 random stratified splits of leave-out 20% cross-validation for evaluation of the top-selected features, and the extra trees classifier was used as the estimator in cross-validation to calculate the prediction score. For regression, we used 20 random splits of leave-out 20% cross-validation and the extra trees regressor as the estimator. The extra trees algorithm was used as the estimator because it generally has consistently good performance without parameter optimization (i.e., the performance is less sensitive to parameter tuning) for datasets with different number of features selected, compared to other models. For all three feature selection methods, the number of features considered was increased incrementally according to the feature orders derived from each selection method. We then selected a cutoff for the optimal set of features for the best selection method based on the value of the cross-validation score, as discussed below in Sec. 3.1.

**2.3 Model selection**

In this work, we performed two separate machine learning tasks related to predicting the stability of perovskite oxides: classification of stable/unstable perovskites and regression of stability based on the $E_{hull}$ values. First, we conducted classification of the $E_{hull}$ values for all 1929 materials to determine which materials are predicted to be unstable ($E_{hull} > 40$ meV/atom above) and which are predicted to be stable ($E_{hull} <= 40$ meV/atom). For classification, we tested five models: logistic regression, support vector machines, decision tree, extra trees classifier and artificial neural network. For regression to predict the $E_{hull}$ values, we tested five regression models: linear regression, kernel ridge regression, decision tree, extra trees regressor and artificial neural network. For both classification and regression, only the features extracted from our feature selection analysis were used to train each model.

For classification, we evaluated and compared the performance of different models with 20 random stratified splits of leave-out 20% cross-validation on the training dataset and compared the average accuracy, precision, recall and $F_1$ score. Accuracy is calculated as the fraction of correctly predicted materials as stable or unstable. Precision is defined as the fraction of predicted stable compounds that are actually stable, and recall is defined as the fraction of actual stable compounds that are predicted stable. To evaluate the overall performance of a model, precision and recall should be considered together. The $F_1$ score is a metric that incorporates both precision and recall, and is calculated as the harmonic mean of precision and recall, i.e., $F_1 = \left( \dfrac{2 \cdot precision \cdot recall}{precision + recall} \right)$.

For regression, we tested each model with 20 random splits of leave-out 20% cross-validation and compared the $R^2$ score, root mean squared error (RMSE) and mean absolute error (MAE).[38] The $R^2$ score is the coefficient of determination, which measures how well the future samples are likely to be predicted by the model. The best possible score for precision, recall and $R^2$ is 1.0. RMSE is the square root of the mean squared (quadratic) error between the predicted value and true value. MAE is the average over the absolute difference between predicted and actual values. For these tests, we compared the results with hyperparameter optimization only performed on included data (the training dataset in the cross validation split) and the results with hyperparameter optimization done on the entire dataset for all classification and regression models, and found there was no significant difference between these two hyperparameter optimization approaches. Therefore, we present all results in this work using the optimized hyperparameters of the classification and regression models on the entire dataset, as this approach is significantly faster. The test performance with hyperparameter optimization only done on included data, and the selected optimized hyperparameters of all classification and regression models are provided in the **DiB**.

**2.4 Performance on various composition subspaces**

As the perovskite dataset is comprised of composition subspaces with various alloying element system, the performance of the trained model can vary based on the frequency of appearance in different composition subspaces. To examine how the performance of the model is influenced by composition subspaces, we removed five sets of perovskite materials from the dataset to serve as five test sets. The perovskite compositions comprising each test set were constructed according to the frequency the constituent elements appeared in the training dataset, as well as the element types present (e.g. alkaline earth versus rare earth elements). This form of model validation is potentially more informative than typical cross-validation, where the training and testing datasets are split randomly. We picked all perovskites that contained certain elements to ensure compounds in the test set have no similar compounds contained in the training set. For each testing set, we trained the model on the dataset excluding the testing set and applied our best model on the excluded data to classify each material as stable/unstable and perform regression to predict $E_{hull}$. We then compared the predicted results with DFT calculated values for verification of the extrapolation performance of our models in various composition subspace.

**2.5 Model validation on completely new test data**

We applied our developed model to 15 new perovskite compounds with element combinations not in the dataset and not considered at any stage of the study until after the final models were determined to predict the $E_{hull}$ values. The 15 new perovskite compounds were divided into three sets according to the frequency the constituent elements appeared in the training dataset. To assess the application of our developed model on these new materials, we computed the $E_{hull}$ values of the new compounds using DFT and compared the DFT-calculated $E_{hull}$ values with the predicted values obtained from our model. While these compounds are in some sense just another test set like those considered in **Sec. 2.4**, they differ in the fact that their DFT values were not available during any stage of the model development, which assures that they have not influenced the model properties in any way.

**3 Results and Discussion**

**3.1 Feature selection**

The total number of features present after removing the constant features was reduced from 962 to 791. We selected the top features using the feature selection methods described in **Sec. 2.2**. **Fig. 2** shows the cross-validation score for each feature selection approach (stability selection, RFE, and univariate feature selection) plotted against the number of features used in the dataset for classification (**Fig. 2a**) and regression (**Fig. 2b**).

For classification, the $F_1$ score increases rapidly with addition of the first 50 features, and converges to a value of about 0.87 after 200 features for stability selection and RFE. For univariate selection, the $F_1$ score increases more slowly with respect to the number of features considered compared to the other two methods and converges to the same $F_1$ score after about 300 features. Among the three feature selection approaches used here, the RFE method results in as high or higher cross-validation $F_1$ score than the other methods and does so with fewer features, making it the optimal approach. We believe univariate selection underperforms RFE because it is based on mutual information that only analyzes the relationship between each individual feature with the target values and fails to capture the relationship between different features. Further, stability selection underperforms RFE because the logistic regression embedded in the stability selection method fails to assign a higher weight for features that correlate non-linearly with the target values. By contrast, RFE uses the extra trees classifier, which has a relatively high accuracy in classification and minimizes over-fitting.[23] Another aspect of RFE which contributes to it yielding the best cross-validation $F_1$ score is that it implements backward feature elimination. Backward feature elimination removes irrelevant features from the beginning of the feature selection process, thus limiting the negative impact irrelevant features may have on the cross-validation score.

To choose the optimal number of features we note that the RFE cross validation $F_1$ score appears to increase little after about 70 features. For example, with 200 features selected by RFE, the cross validation $F_1$ score converges to the maximum 0.88, and the cross-validation accuracy, precision and recall are 0.93, 0.89, 0.86 respectively. However, with 70 features selected by RFE, the cross-validation $F_1$ score is 0.87, and the cross-validation accuracy, precision and recall are 0.93, 0.89, 0.85, respectively. The change in these model metrics is very small between 70 and 200 features, and when fewer than 70 features are used, the $F_1$ score drops at a higher rate compared to the region from 70 to 200 features. We therefore decided to only use the top 70 features to fit our machine learning models. Using the fewest number of features possible without significant loss of accuracy gives the best chance to avoid over-fitting and construct the simplest and most accurate predictive model.

For regression, the fitting performance was evaluated by the average $R^2$ score of 20 random splits of leave out 20% cross-validation. We again find that the RFE method has as high or higher a figure of merit (here $R^2$) than the other methods and achieves this performance with fewer features, making it the optimal approach. We also again find that the figure of merit changes little after about 70 features. Similar to the feature selection analysis for classification discussed above, the $R^2$ score reached its maximum value of 0.89 when 200 features were used, and $R^2$ still maintained a value of 0.89 when 70 features were used. The change in $R^2$ is very small between 70 and 200 features, and when fewer than 70 features are used, the $R^2$ score drops at a higher rate compared to the region from 70 to 200 features. Thus, we selected the top 70 features for further analysis. A complete list of top 70 features selected for classification and regression is also provided in the **DiB**.

To know the important chemical and physical elemental properties that closely relate to thermodynamic stability, we counted the number of appearances of the elemental properties among the sets selected by both classification and regression of top 70 features (for example, if one feature is the average modulus of elasticity and another feature is the difference of modulus of elasticity between A-site atoms and B-site atoms, we count the elemental property "modulus of elasticity" as appearing twice on the lists). **Table 1** shows the list of frequently occurring (occurring >= 5 times on both lists) elemental properties along with the number of times they appear. The frequent appearance of chemical and physical properties may not identify all important elemental properties, but it is reasonable to expect that many of the critical properties will appear multiple times, indicating they can influence the stability in multiple ways (e.g. A-site atom properties, B-site atom properties, or the relative value between A-site and B-site atom properties). The ability to impact stability of most properties that occur frequently can be explained in terms of physical or chemical mechanisms. For example, the number of unfilled valence orbitals can influence the stable oxidation states available for the atoms in the perovskite, thus impacting the overall stability of the structure as the perovskite favors certain oxidation states. The influence of BCC energy (the energy of the element in the Open Quantum Materials Database (OQMD)[39] ground state crystal structures minus that of the BCC DFT value, all at 0 K) is less obvious, but, for example, may provide information about the bonding tendencies of elements that impact their stability in the perovskite structure. Perovskite materials often contain cations in the 3+ oxidation state on A- and B- sites, so the third ionization potential is expected to influence the stability of compounds by indicating how easily an atom can form a cation with an oxidation state of 3+. In addition, there are properties (HHIp and HHIr) that appear frequently and are related to the Herfindahl−Hirschman Index (HHI)[40] of elements, which are hard to explain in terms of fundamental physics and chemistry. HHI is related to elemental reserves and production, and there may be an indirect correlation between factors setting the HHI (e.g, scarcity of elements in the Earth's crust) and physical and chemical properties relating to stability.

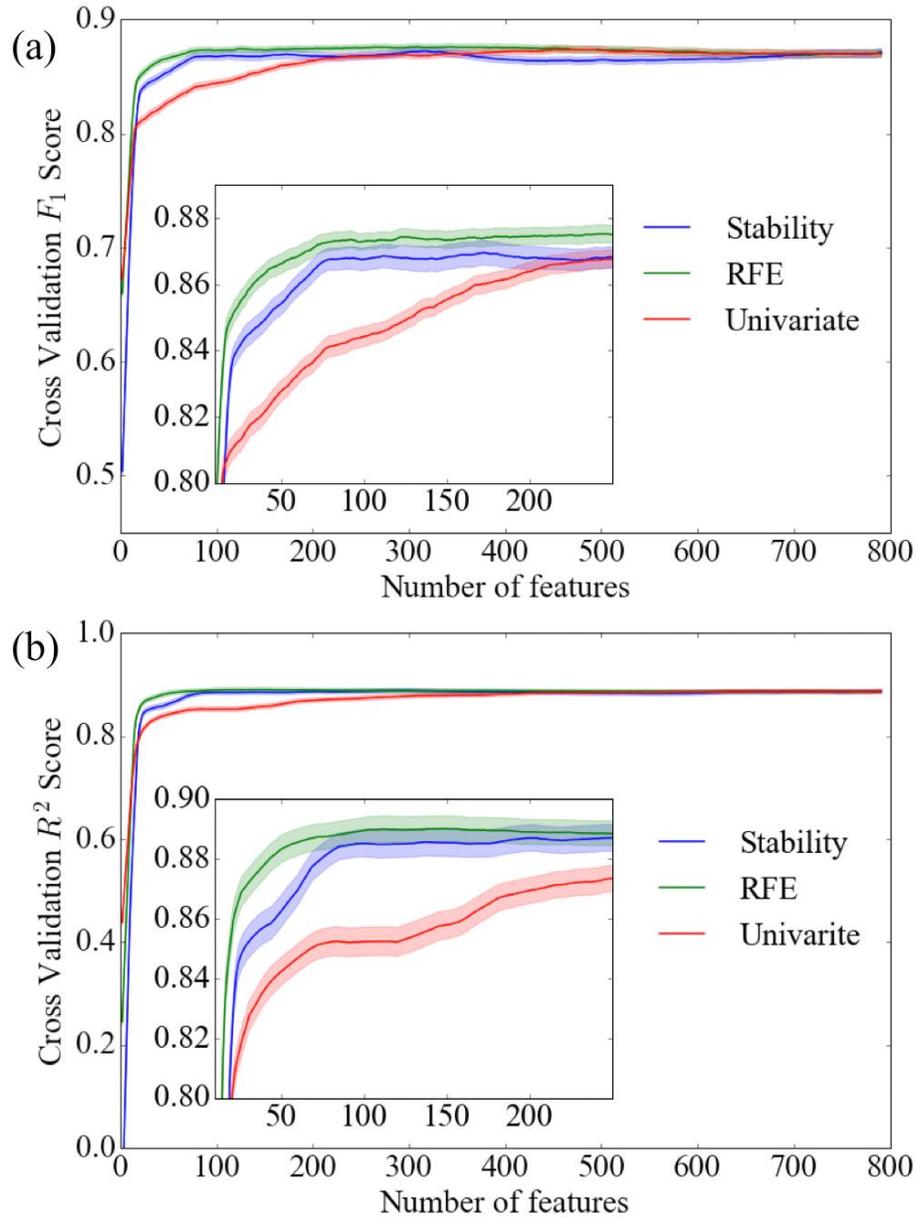

**Fig. 2** (a) Value of the cross-validation $F_1$ score with increasing number of features for three selection methods used in classification: stability selection, recursive feature elimination (RFE) and univariate selection based on mutual information. The $F_1$ score is calculated as the average of 20 random stratified splits of leave out 20% cross-validation. The $F_1$ score for stability selection decreases after 200 features because more irrelevant features are included, resulting in some over-fitting of the training data. (b) Value of the cross-validation $R^2$ score with increasing number of features for three selection methods used in regression: stability selection, recursive feature elimination (RFE) and univariate selection based on mutual information. The $R^2$ score is calculated as the average of 20 random splits of leave out 20% cross-validation.

**Table 1**. A list of frequent chemical and physical elemental properties in selected features. The value in parentheses after each property indicates the number of times that property appears in the combined top 70 features of both the extra trees classification and extra trees regression model (see text for details). The detailed description of each elemental property is listed in Table S4 of the supplemental information in **DiB**. Each column lists the feature modifiers that, when applied to the property given as the column header, generates the relevant feature. Feature modifiers include: "AB_avg" – average of all A- and B-site atoms, "w_avg" –weighted average by atom fraction, "H" –atom of highest atom fraction, "Asite" – A-site atoms, "Bsite" – B-site atoms, "min" – minimum value among all atoms, "max" – maximum value among all atoms, "AB_diff" – difference of A- and B-site atoms, "AB_ratio" – ratio of A- and B-site atoms, "range" – range among all atoms. "(2)" after a feature modifier means the feature appears in both sets.

| **Elemental Property** | Number of unfilled valence orbitals (9) | Coefficient of thermal expansion (8) | HHIp (8) | Mendeleev Number (7) | BCC energy (7) | Group number (6) |
|---|---|---|---|---|---|---|
| **Feature Modifiers** | Bsite_min(2), Bsite_w_avg(2), Asite_min, Asite_max(2), Asite_w_avg, H_Bsite | H_Bsite, AB_avg(2), Asite_max, Asite_w_avg, Bsite_w_avg(2), AB_ratio | Asite_min(2), Bsite_max, Asite_w_avg, AB_avg(2), AB_diff, Bsite_max | Bsite_max, Bsite_w_avg, Asite_min, AB_avg, AB_diff, AB_ratio(2) | Asite_min(2), Bsite_range, AB_diff(2), AB_avg, Bsite_range | Bsite__avg, Biste_max(2), Bsite_w_avg(2), H_Bsite |
| **Elemental Property** | BCC Fermi (6) | Number of d valance orbitals (6) | Heat of Vaporization (5) | HHIr (5) | Third ionization potential (5) | |
| **Feature Modifiers** | Bsite_w_avg, AB_avg(2), AB_ratio(2), AB_diff | Bsite_max (2), Bsite_w_avg(2), H_Bsite, Bsite_min | Asite_min, Bsite_w_avg, AB_avg(2), AB_ratio | H_Bsite, Asite_min, Asite_w_avg, AB_avg, AB_diff | AB_avg, Bsite_range, Asite_max, Bsite_max, AB_diff | |

### 3.2 Classification of Stable/Unstable perovskites

We tested five classifiers, including: logistic regression, support vector machines, decision tree, neural network and extra trees. These classifiers were fit using the training dataset with the best 70 features obtained from feature selection (see **Sec. 3.1**). The parameters of the five models were optimized based on the cross-validation $F_1$ score. **Table 2** summarizes the performance of these five models for accuracy, precision, recall and $F_1$ score, the values of which were averaged from 20 random stratified splits of leave-out 20% cross-validation. From **Table 2**, the support vector machines, neural network and extra trees classifier models have an $F_1$ score over 0.87 and accuracy over 0.92. We also found that the classification performance is much less sensitive to parameter tuning for the extra trees classifier compared to SVM and neural network classifier (data not shown), which is an additional advantage of the extra trees approach.

**Table 2** Comparison of classification accuracy, precision, recall and $F_1$ score between five classifiers. In (score +/- error), score and error are calculated as the mean score, and 2×standard deviation of 20 random stratified splits of leave out 20% cross-validation, respectively.

| Model | Logistic Regression | SVM with RBF kernel | Decision Tree | Neural Network | Extra Trees |
|---|---|---|---|---|---|
| Accuracy | 0.81 (+/- 0.03) | 0.93 (+/- 0.02) | 0.88 (+/- 0.03) | 0.93 (+/- 0.02) | 0.93 (+/- 0.02) |
| Precision | 0.63 (+/- 0.05) | 0.89 (+/- 0.05) | 0.81 (+/- 0.037) | 0.88 (+/- 0.04) | 0.89 (+/- 0.07) |
| Recall | 0.87 (+/- 0.07) | 0.86 (+/- 0.05) | 0.79 (+/- 0.08) | 0.86 (+/- 0.05) | 0.87 (+/- 0.05) |
| $F_1$ score | 0.73 (+/- 0.04) | 0.88 (+/- 0.04) | 0.80 (+/- 0.07) | 0.87 (+/- 0.03) | 0.88 (+/- 0.03) |

The performance of three classification models: support vector machines, neural network and extra trees classifier was analyzed using the receiver-operator characteristic curve (ROC curve, see **Fig. 3**). We did not consider the ROC of the logistic regression and decision tree due to its inferior performance based on the metrics shown in **Table 2**. The ROC curve can effectively illustrate the performance of a binary classifier system as its discrimination threshold is varied.[41] For each model, we split the dataset into 10 random stratified folds and iteratively predicted the probability of being stable in each fold based on the other 9 folds (10-fold cross-validation). We then calculated the true positive rate (sensitivity) and false positive rate (1 - specificity) at various probabilistic confidence threshold applied on the predicted probability of being stable of all the compounds. The true positive rate is the fraction of true stable compounds identified successfully by the model, and the false positive rate is the fraction of unstable compounds mistakenly identified as stable compounds by the model. The points on the ROC curve show true positive rate and false positive rate scores at various probability thresholds. **Fig. 3** shows a representative ROC curve with the x-axis as the true positive rate and the y-axis as the false positive rate. The area under the ROC curve (Area Under Curve, or AUC) is a measure of the overall performance and is interpreted as the average value of sensitivity for all possible values of specificity. The closer AUC is to 1, the better the overall prediction performance of the model. The ROC curve in **Fig. 3** also offers guidance in selecting the appropriate probabilistic confidence threshold for using our model to evaluate whether a given perovskite oxide composition is stable or unstable, given a desired value for the resulting precision and recall scores. The AUC for SVM, neural network and extra trees classifiers are 0.974, 0.976 and 0.980, respectively. The higher AUC value for the extra trees classifier further demonstrates it is the best model for the prediction of whether a given perovskite is stable or unstable.

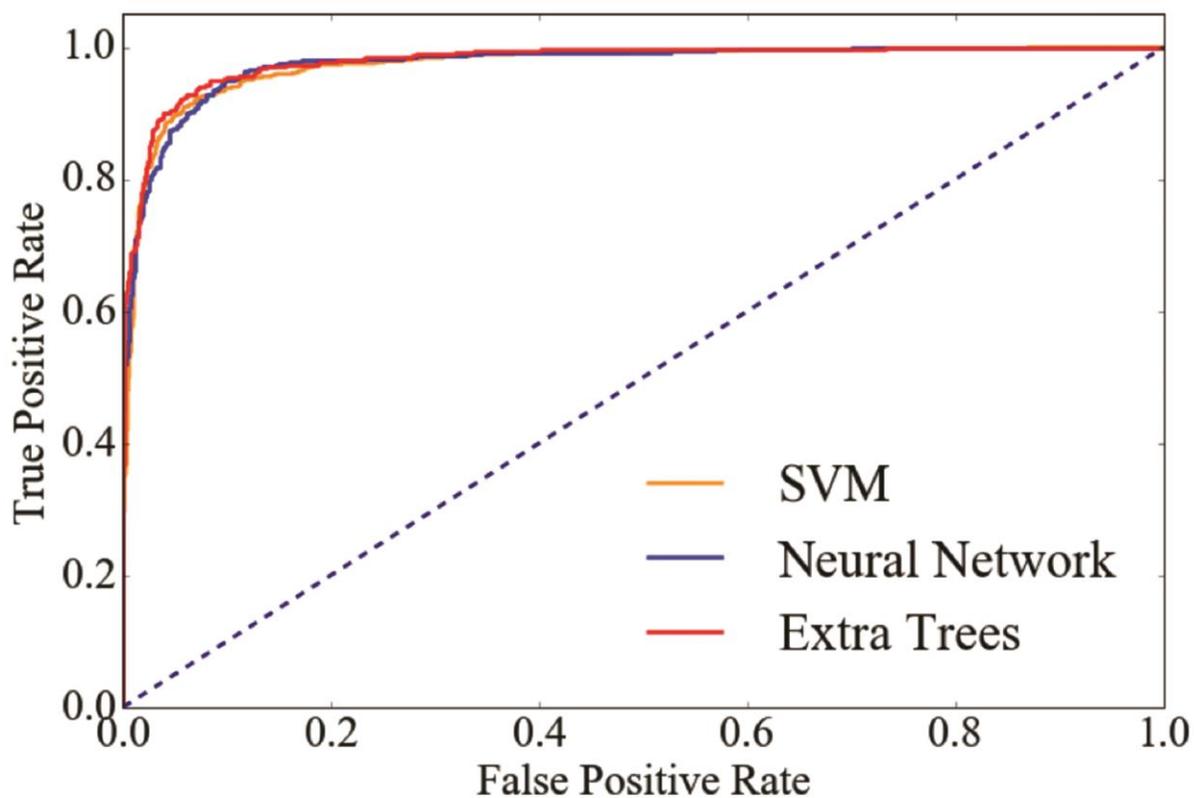

**Fig. 3** Receiver-operator characteristic (ROC) curve of SVM, neural network and extra trees classifier. The area under the ROC curve (AUC) for SVM, neural network and extra trees classifiers are 0.974, 0.976 and 0.980, respectively.

### 3.3 Regression of $E_{hull}$ of perovskites

In addition to simply classifying whether a perovskite oxide is expected to be stable or unstable, it is informative to predict a numerical value for $E_{hull}$. Therefore, we also trained five regression models to predict the $E_{hull}$ values, and the best model was selected to perform additional model validation (see Sec. 3.4). In addition to performing regression on the $E_{hull}$ values, we also performed an analogous analysis for regression of the DFT-calculated formation energies, measured relative to the DFT-energies of the appropriate concentrations of elemental end-member compounds. Because the formation energy values are not a direct indicator of material stability they are somewhat tangential to the focus of this study, therefore the details, data and analysis of regression of formation energies can be found in the **DiB**. The dataset we used for regression contained a total number of 1918 compounds and their DFT calculated $E_{hull}$ values. 11 compounds with $E_{hull}$ values greater than 400 meV/atom were removed from the original dataset of 1929 perovskites, as our main interest was to get high prediction accuracy for stable and nearly stable compounds, and a small number of highly unstable materials could have a negative impact on the model accuracy in the region of interest. We used the feature selection approach described in **Sec. 3.2** to generate the complete set of model features. We selected the top 70 features for $E_{hull}$ regression from a total of 791 features.

We evaluated and compared the training dataset cross-validation $R^2$ score, RMSE and MAE values of five regression models: linear regression, kernel ridge regression, decision tree regressor, extra trees regressor and artificial neural network. **Table 3** provides a summary of the performance for each model via their $R^2$ score, RMSE and MAE values on the dataset. The $R^2$, RMSE and MAE scores were calculated as the average value of 20 runs (sometimes called splits) of leave-out 20% cross-validation. The kernel ridge regression with a radial basis function (rbf, also called Gaussian) kernel and the extra trees regressor perform very closely, which both have a cross-validation $R^2$ score of about 0.89, RMSE of less than 30 meV/atom, and MAE of less than 17 meV/atom in the $E_{hull}$ values. We note here that the RMSE value obtained for the kernel ridge regression model of 28.5 (+/- 7.5) meV/atom is on the same order as typical DFT errors for calculation of $E_{hull}$ relative to elemental reference states when compared to experiments (as described in **Sec. 2.1**), [7, 27] suggesting that significantly greater accuracy would not be useful and the current model may be sufficiently accurate to use in place of full DFT calculations.

**Table 3** Comparison of $R^2$, RMSE and MAE values between five regression models for prediction of $E_{hull}$. The values are listed as (score +/- error), where score and error are calculated as the mean value, and 2×standard deviation of 20 random splits of leave out 20% cross-validation, respectively.

| Model | | Linear Regression | Kernel Ridge with RBF | Decision Tree | Neural Network | Extra Trees |
|---|---|---|---|---|---|---|
| **Model Evaluation Metrics** | $R^2$ | 0.725 (+/- 0.076) | 0.894 (+/- 0.059) | 0.737 (+/- 0.102) | 0.874 (+/- 0.063) | 0.888 (+/- 0.054) |
| | RMSE (meV/atom) | 46.1 (+/- 6.1) | 28.5 (+/- 7.5) | 44.3 (+/- 11.1) | 31.4 (+/- 7.5) | 29.4 (+/- 7.3) |
| | MAE (meV/atom) | 32.7 (+/- 2.8) | 16.7 (+/- 2.3) | 26.1 (+/- 4.8) | 18.7 (+/- 2.1) | 16.0 (+/- 2.2) |

We split the dataset into 10 random folds in the same way as the classification work, and predicted the $E_{hull}$ values in each fold based on the other 9 folds using kernel ridge regression (10-fold cross validation). **Fig. 4** is a representative plot of the 1918 predicted $E_{hull}$ using kernel ridge regression versus the DFT calculated values. The predicted values do vary somewhat over different cross-validation tests due to different random splitting, but there is no significant difference in the overall residuals distribution. The plot of residuals as an inset in **Fig. 4** shows most compounds (85%) were predicted within the RMSE error of 28 meV/atom (average RMSE error given by the best model).

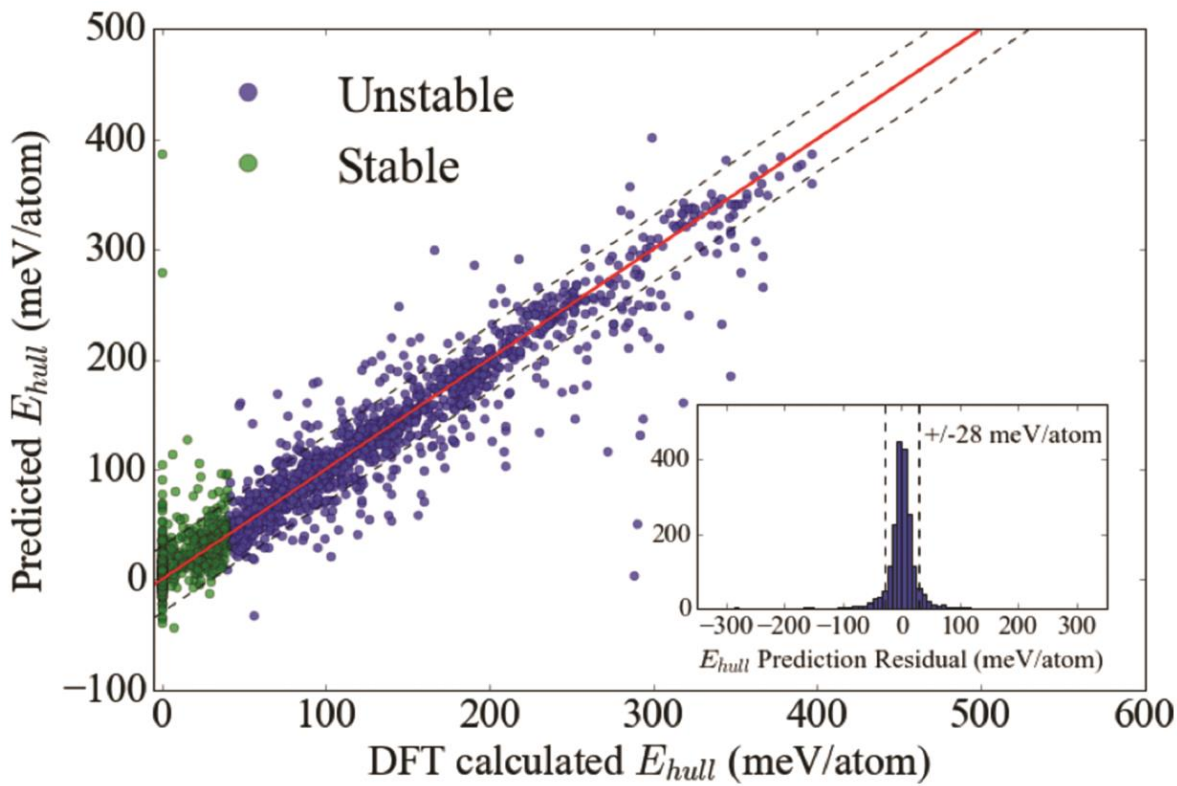

**Fig. 4** Fitted $E_{hull}$ values versus the DFT calculated values. The green and blue points represent materials that were found to be stable and unstable based on DFT calculations, respectively. The y=x line is shown as a solid red line and the surrounding dashed lines represent +/- the 28 meV/atom RMSE (in **Table 3**) shifts from the y=x line. The inset residual plot shows the histogram of prediction error between predicted $E_{hull}$ and DFT calculated $E_{hull}$.

### 3.4 Performance on various composition subspaces

In this section, we analyzed the validity of our model for different perovskite composition subsets. We did this by removing different material subsets from the full dataset to use as the testing data, and trained the extra trees classifier and kernel ridge regression model with all remaining perovskite materials. **Fig. 5** shows the frequency heat map of the elements sampled on the A- and B-site for all perovskites in the dataset. Because the perovskite compositions in the full dataset are unevenly sampled, we chose five test sets, where each set represents a composition subspace based on the frequency the constituent elements appeared in the training dataset or a particular class of elements, e.g., alkaline earth versus rare earth elements on the A-site sublattice. Here, we describe the five different material subsets used for extrapolation testing, using the perovskite structure notation of $(\{A\},\{A'\},\{A''\})(\{B\},\{B'\},\{B''\})O_3$. With this notation, the parentheses are used to separate elements alloyed on the A-site and those alloyed on the B-site, while the curly braces are used to denote a set of elements that are alloyed on a particular site. The $\{A\}$, $\{A'\}$ and $\{A''\}$ represent lists of different A-site elements sampled in the dataset, while the $\{B\}$, $\{B'\}$, and $\{B''\}$ represent lists of different B-site elements sampled in the dataset. Note that a list may contain 0, 1 or more elements. For example, the material $SrFeO_3$ would consist of $\{A\}$=Sr, $\{A'\}$=$\{A''\}$=0 (no elements), and $\{B\}$=Fe, $\{B'\}$=$\{B''\}$=0 (no elements). As a second example, the material $(\{Ba, Sr\},\{La\})(\{Co,Fe\})O_3$ indicates the set of materials where $\{A\}$=Ba or Sr, $\{A'\}$=La, $\{A''\}$=0 and $\{B\}$=Co or Fe, $\{B'\}$=$\{B''\}$=0. Note that we only consider up to 3 species on each sublattice in this dataset, so listing $\{A\},\{A'\},\{A''\}$ covers all possible A-site combinations in our data, and analogously for B-sites. The five sets considered here are: 1. $(\{Ba,Ca\})(\{B\},\{B'\},\{B''\})O_3$, 2. $(\{Pr,Dy,Gd,Ho\},\{Pr,Dy,Gd,Ho\})(\{B\},\{B'\},\{B''\})O_3$, 3. $(\{Ba,Sr\})(\{Fe\},\{B'\},\{B''\})O_3$, 4. $(\{A\},\{A'\},\{A''\})(\{V,Cr,Ti,Ga,Sc\},\{V,Cr,Ti,Ga,Sc\})O_3$, and 5. $(\{Bi,Cd,Mg,Ce,Er\},\{A'\},\{A''\})(\{B\},\{B'\},\{B''\})O_3$. The description of each set and the rationale for its construction is provided below:

1. $(\{Ba,Ca\})(\{B\},\{B'\},\{B''\})O_3$: this set only contains the alkaline earth elements Ba and Ca on the A-site (i.e., mixed Ba-Ca materials were considered), and the B, B' and B" are any sampled B-site element or combinations thereof. This set was chosen to examine the model validity when the majority of materials containing alkaline earth elements on the A-site were removed.

2. $(\{Pr,Dy,Gd,Ho\},\{Pr,Dy,Gd,Ho\})(\{B\},\{B'\},\{B''\})O_3$: this set only contains the rare earth elements Pr, Dy, Gd, and Ho on the A-site, and the B, B' and B" are any sampled B-site element or combinations thereof. This set was chosen to examine the model validity when the majority of materials containing rare earth elements on the A-site were removed.

3. $(\{Ba,Sr\})(\{Fe\},\{B'\},\{B''\})O_3$: this set contains all compounds with only the alkaline earth elements Ba and Sr on the A-site and containing Fe (plus other elements) on the B-site. This set was chosen because it represents the set containing the most frequently sampled elements in the full dataset.

4. $(\{A\},\{A'\},\{A''\})(\{V,Cr,Ti,Ga,Sc\}\{V,Cr,Ti,Ga,Sc\})O_3$: this set contains all compounds with only V, Cr, Ti, Ga and Sc atoms on the B-site and any element on the A-site. This set was chosen because it represents the set containing moderately sampled elements from the full dataset. This dataset also represents a set of compounds determined by their B-site elements, as opposed to set (1)-(3) and (5), which are determined solely by A-site elements.

5. ({Bi,Cd,Mg,Ce,Er},{A'},{A"})({B},{B'},{B"})$O_3$: this set contains all compounds with one element from the set of Bi, Cd, Mg, Ce, Er plus other elements on the A-site and any element on the B-site. This set was chosen because it represents the set with the least frequently sampled elements from the full dataset.

The number of compounds in the five data sets are (1) 50, (2) 54, (3) 53, (4) 47 and (5) 40. For each set, we predicted the $E_{hull}$ values using this manually targeted test scheme, i.e., we selected our best models (extra trees model for classification and kernel ridge regression model for regression) with the hyperparameters determined in **Sec. 2.3** without re-optimizing the model for the manually split set here, and we trained the models on the training dataset with the set of testing compounds removed and predicted the $E_{hull}$ values and classification results for the testing compounds in each composition set.

| B Site | Ba | Sr | La | Y | Pr | Ca | Zn | Dy | Gd | Ho | Nd | Sm | Bi | Cd | Sn | Mg | Ce | Er |
|---|---|---|---|---|---|---|---|---|---|---|---|---|---|---|---|---|---|---|
| Fe | 286 | 106 | 104 | 82 | 81 | 69 | 40 | 11 | 11 | 11 | 11 | 11 | 9 | 6 | 6 | 5 | 2 | |
| Mn | 138 | 129 | 106 | 98 | 101 | 112 | 40 | 7 | 7 | 7 | 7 | 7 | | | | | 1 | 1 |
| Co | 126 | 96 | 106 | 101 | 103 | 73 | 40 | 6 | 7 | 6 | 6 | 7 | 6 | | | | | |
| Ni | 114 | 90 | 103 | 101 | 104 | 70 | 40 | 6 | 5 | 6 | 6 | 6 | | 3 | 3 | | | |
| V | 10 | 123 | 36 | 37 | 37 | 10 | | 9 | 9 | 9 | 8 | 9 | | | | 3 | 2 | |
| Cr | 7 | 12 | 52 | 52 | 52 | 7 | | 5 | 6 | 6 | 6 | 6 | | | | | | |
| Ti | 11 | 14 | 38 | 39 | 39 | 7 | | 6 | 6 | 6 | 5 | 6 | | | | | | |
| Ga | 16 | 14 | 34 | 35 | 35 | 6 | | 6 | 6 | 6 | 6 | 6 | | | | | | |
| Sc | 6 | 9 | 35 | 35 | 35 | 6 | | 6 | 6 | 6 | 6 | 6 | | | | | | |
| Zr | 68 | 7 | 4 | 4 | 13 | | | | | | | | | 3 | 6 | 6 | | |
| Mg | | 3 | 27 | 27 | 27 | | | | | | | | | | | | | |
| Hf | 34 | 7 | 4 | 4 | 13 | | | | | | | | | | | | | |
| Nb | 29 | 7 | 4 | 4 | 12 | | | | | | | | | | | | | |
| Ta | 28 | 7 | 4 | 4 | 13 | | | | | | | | | | | | | |
| Al | 10 | 8 | | | | | | | | | | | | | | | | |
| Cu | 14 | 4 | | | | | | | | | | | | | | | | |
| Zn | 13 | 4 | | | | | | | | | | | | | | | | |
| Mo | 7 | 3 | | | | | | | | | | | | | | | | |
| Ir | 4 | 3 | | | | | | | | | | | | | | | | |
| Os | 4 | 3 | | | | | | | | | | | | | | | | |
| Pd | 4 | 3 | | | | | | | | | | | | | | | | |
| Pt | 4 | 3 | | | | | | | | | | | | | | | | |
| Re | 4 | 3 | | | | | | | | | | | | | | | | |
| Rh | 4 | 3 | | | | | | | | | | | | | | | | |
| Ru | 4 | 3 | | | | | | | | | | | | | | | | |
| Y | 4 | 3 | | | | | | | | | | | | | | | | |
| Tc | 4 | | | | | | | | | | | | | | | | | |
| Ge | | 3 | | | | | | | | | | | | | | | | |
| Si | | 3 | | | | | | | | | | | | | | | | |
| Sn | | 3 | | | | | | | | | | | | | | | | |
| W | 1 | | | | | | | | | | | | | | | | | |

A Site

**Fig. 5** Frequency heat map of all constituent elements in the dataset of 1929 perovskites. The value in each square block indicates the number of compounds with the corresponding elements in A- and B- sites appeared in the dataset.

**Fig. 6** shows the predicted $E_{hull}$ values versus the DFT calculated results for each perovskite subset using our manually targeted cross-validation scheme. **Table 4** shows the confusion matrix of the classification results and $R^2$ score, RMSE and MAE values of the regression results.

For the case of regression (**Fig. 6**), the data set (1), (2), (3), (4) all performed well, with predicted RMSE values of 26.4, 16.3, 25.1 and 28.3 meV/atom, respectively, which are similar to the overall average errors in the fit (see **Table 3**). Overall, the set (2) performed best as it has the smallest RMSE compared to other sets and also has a high $R^2$ value of 0.851. Importantly, the high accuracy of regression on the set (1), (2) and (3) makes this model very useful in predicting the stability of perovskite oxides in the composition space relevant for high activity SOFC cathodes[42, 43]. By comparison, the set (5) performs noticeably worse with a predicted RMSE of 72.7 meV/atom. This worse performance of set (5) is not surprising. Sets (1), (2), (3) most likely performed well due to the large number of compounds in the training set containing the same A-site elements. Correspondingly, set (5) displayed worse performance due to there being few compounds in the training set containing the same A-site elements.

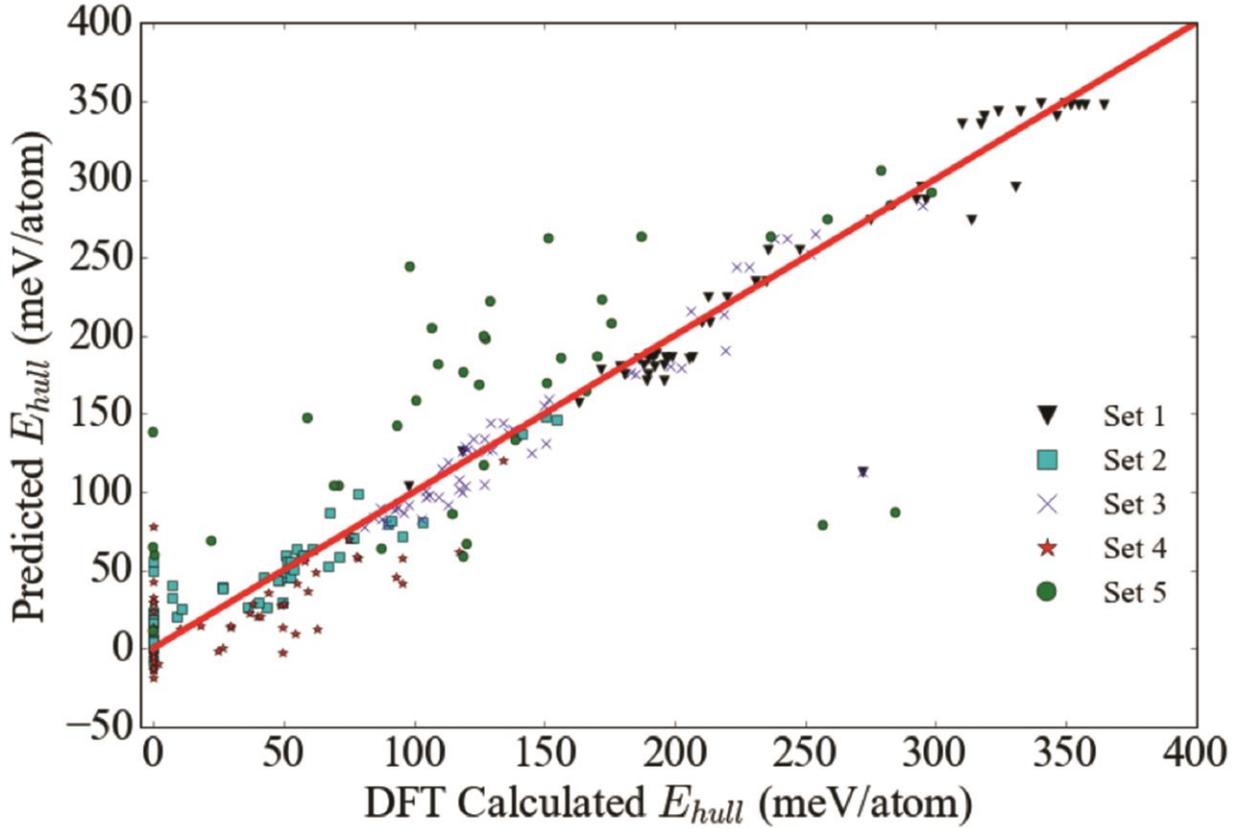

**Fig. 6** Plot of predicted $E_{hull}$ values using the kernel ridge regression model, versus the DFT calculated results for five different perovskite subsets. The RMSE values of $E_{hull}$ for Set 1 (\{Ba,Ca\})(\{B\},\{B'\},\{B"\})$O_3$, Set 2 (\{Pr,Dy,Gd,Ho\},\{Pr,Dy,Gd,Ho\})(\{B\},\{B'\},\{B"\})$O_3$, Set 3 (\{Ba,Sr\})(\{Fe\},\{B'\},\{B"\})$O_3$, Set 4 (\{A\},\{A'\},\{A"\})(\{V,Cr,Ti,Ga,Sc\}\{V,Cr,Ti,Ga,Sc\})$O_3$, and Set 5 (\{Bi,Cd,Mg,Ce,Er\},\{A'\},\{A"\})(\{B\},\{B'\},\{B"\})$O_3$ are (1) 26.4, (2) 16.3, (3) 25.1, (4) 28.3 and (5) 72.7 meV/atom, respectively.

For the case of classification (**Table 4**), set (5) displayed excellent classification results, with 95% correct predictions. This is somewhat surprising given the relatively poor performance for data set (5) in the regression results discussed above (**Fig. 6**). However, 4 out of the 5 stable compounds in set (5) have an $E_{hull}$ of 0 meV/atom, and most of the unstable compounds have very high $E_{hull}$ values. Thus, the classification task is relatively straightforward for set (5). By comparison, set (4) contains a number of unstable compounds with $E_{hull}$ values close to the stable/unstable cutoff value of 40 meV/atom, which makes classifying these compounds as stable or unstable more challenging. This set therefore performs the most poorly of the sets (1)-(5) in classification, although it is as good as any of the other sets (1)-(4) in regression. In general, our model shows good classification performance in these targeted cross-validation cases, even when the number of compounds of a particular set is limited. Furthermore, comparison across the regression and classification accuracy shows that their trends can be somewhat different, depending in particular on how close the compounds are to the classification boundary.

**Table 4** Comparison of confusion matrix, $R^2$, RMSE and MAE values for the five perovskite composition subsets. In the confusion matrix, true unstable (stable) is the number of unstable compounds that are successfully predicted as unstable (stable), and false unstable (stable) is the number of unstable compounds that are mistakenly predicted as stable (unstable).

| Stability Prediction | Stability Classification | | Energy above convex hull | | |
|---|---|---|---|---|---|
| Prediction Score | True Unstable | False Stable | $R^2$ | RMSE (meV/atom) | MAE (meV/atom) |
| | False Unstable | True Stable | | | |
| (\{Ba,Ca\})(\{B\},\{B'\},\{B"\})O$_3$ | 50 | 0 | 0.854 | 26.4 | 13.4 |
| | 0 | 0 | | | |
| (\{Pr,Dy,Gd,Ho\},\{Pr,Dy,Gd,Ho\})(\{B\},\{B'\},\{B"\})O$_3$ | 26 | 5 | 0.851 | 16.3 | 12.0 |
| | 1 | 22 | | | |
| (\{Ba,Sr\})(\{Fe\},\{B'\},\{B"\})O$_3$ | 53 | 0 | 0.795 | 25.1 | 13.2 |
| | 0 | 0 | | | |
| (\{A\},\{A'\},\{A"\})(\{V,Cr,Ti,Ga,Sc\}\{V,Cr,Ti,Ga,Sc\})O$_3$ | 14 | 6 | 0.371 | 28.3 | 22.8 |
| | 4 | 23 | | | |
| (\{Bi,Cd,Mg,Ce,Er\},\{A'\},\{A"\})(\{B\},\{B'\},\{B"\})O$_3$ | 34 | 1 | 0.345 | 72.7 | 56.4 |
| | 1 | 4 | | | |

### 3.5. Model Application: new compounds

As a further test of our model validation, we manually generated 15 new perovskite compounds and applied our extra trees classification and kernel ridge regression models to predict the stability of these new compounds. After predicting whether each new material would be stable/unstable and its predicted $E_{hull}$ value, we then performed DFT calculations to obtain the calculated $E_{hull}$ values for comparison with our predictions. In an analogous manner as was done in **Sec. 3.4**, we constructed the new compounds based on the frequency the constituent elements appeared in the dataset, and divided them into three sets (set A, B and C) by the order of frequency. The precise compositions of the new compounds do not appear in the original dataset. The compounds in set A represent the perovskites consisting of elements very frequently sampled in the dataset. For set A, we expect the models can learn the most information about materials similar to the compounds in set A from the training dataset. The compounds in set B represent the group of compounds with elements less frequently sampled in the dataset. For set B, we expect the models will obtain limited information about materials similar to the compounds in set B from the training dataset. The compounds in set C represent the group of perovskites with element combinations not sampled at all in the training dataset. For set C, we expect the models will obtain almost zero information about materials similar to those in set C. While these tests are similar to those performed in **Sec. 3.4** they are distinct because the DFT energies of these compounds were never seen during any stage of the fitting process, and only made available after the project was complete. They therefore provide a particularly demanding test of our machine learning models.

**Table 5** shows the prediction results of all 15 new perovskites and a comparison with DFT calculated results. From the results shown in **Table 5**, 3 out of 5 compounds in set A get correct stable/unstable classification and 4 out of 5 compounds in set A get prediction error within 25 meV/atom. The mean absolute error of $E_{hull}$ prediction is 46.7 meV/atom because of a large prediction error on one compound. Set B has all correct classification of stability since all of them are unstable compounds, and 2 out of 5 compounds in set B get prediction error within 25 meV/atom, and the mean absolute error is 98.7 meV/atom. In set C, 3 out of 5 compounds are classified correctly, while only 1 out of 5 compounds in set C get prediction error within 25 meV/atom and the mean absolute error of $E_{hull}$ prediction is 166.0 meV/atom. These tests demonstrate that predicting completely new perovskites is improved when more information of the constituent elements is contained in the training dataset. For example, all elements in the compound $BaFe_{0.25}V_{0.75}O_3$ are frequent in the dataset, and the prediction error of $BaFe_{0.25}V_{0.75}O_3$ is only 8.5 meV/atom. In contrast, our worst prediction is for $La_{0.25}Pr_{0.75}Ge_{0.5}Sn_{0.5}O_3$ in set C, which contains Ge and Sn which both only appeared 3 times in the dataset of more than 1900 perovskites and only appeared together when Sr was on the A-site. The prediction error of La$_{0.25}$Pr$_{0.75}$Ge$_{0.5}$Sn$_{0.5}$O$_3$ is 472.5 meV/atom, due to their being very limited information about Ge and Sn in the training dataset. It is worth noting that the prediction of all these compounds by the trained models only takes 0.0019 seconds on a machine with one 2-core 2.9 GHz Intel Core i5 CPU, while the computation time using DFT calculation takes approximately 450,000 seconds (8-9 hours for one compound) on 24 2.2 GHz processors, which would be approximately more than 10$^7$ times slower than the machine learning approach, assuming linear time scaling with number of cores and clock speed. These results suggest the model is useful to quickly identify stable and near-stable perovskite oxides when the elements appear frequently in the database, but should probably not be used for elements that are infrequent or not represented in the database.

**Table 5** Comparison of DFT calculated E$_{hull}$ and model-predicted E$_{hull}$ for the three perovskite composition subsets. Stable/unstable classification is predicted by extra trees classifier, and E$_{hull}$ is predicted by kernel ridge regression.

| | Composition | DFT calculated E$_{hull}$ (meV/atom) | Predicted stability | Predicted E$_{hull}$ (meV/atom) | Difference in E$_{hull}$ (meV/atom) |
|---|---|---|---|---|---|
| Set A | BaFe$_{0.25}$V$_{0.75}$O$_3$ | 111.6 | unstable | 120.1 | 8.5 |
| | La$_{0.5}$Y$_{0.5}$Co$_{0.5}$Mn$_{0.5}$O$_3$ | 192.4 | stable | 23.6 | 168.8 |
| | PrTi$_{0.75}$V$_{0.25}$O$_3$ | 33.8 | unstable | 56.6 | 22.8 |
| | SrCr$_{0.5}$Mn$_{0.5}$O$_3$ | 64.8 | unstable | 82.5 | 17.7 |
| | Y$_{0.75}$Sr$_{0.25}$VO$_3$ | 16.6 | stable | 32.2 | 15.6 |
| Set B | Ca$_{0.5}$Y$_{0.5}$Ni$_{0.5}$Sc$_{0.5}$O$_3$ | 293.0 | unstable | 9.4 | 283.6 |
| | CaMg$_{0.25}$Ti$_{0.75}$O$_3$ | 110.2 | unstable | 88.8 | 21.4 |
| | Dy$_{0.75}$Ba$_{0.25}$Ga$_{0.5}$Mg$_{0.5}$O$_3$ | 431.7 | unstable | 283.2 | 148.5 |
| | Gd$_{0.25}$Sr$_{0.75}$NiO$_3$ | 190.7 | unstable | 158.2 | 32.5 |
| | Y$_{0.75}$Ca$_{0.25}$Co$_{0.75}$Cr$_{0.25}$O$_3$ | 96.1 | unstable | 88.8 | 7.3 |
| Set C | BiPt$_{0.5}$Pd$_{0.5}$O$_3$ | 96.6 | unstable | 37 | 59.6 |
| | CeReO$_3$ | 390.2 | stable | 155.5 | 234.7 |
| | $Dy_{0.5}Zn_{0.5}Al_{0.5}Zr_{0.5}O_3$ | 137.9 | unstable | 179.3 | 41.4 |
| | $Dy_{0.75}Nd_{0.25}RuO_3$ | 89.7 | stable | 111.3 | 21.6 |
| | $La_{0.25}Pr_{0.75}Ge_{0.5}Sn_{0.5}O_3$ | 496.7 | unstable | 24.2 | 472.5 |

## 4 Conclusion

In this work, we used machine learning algorithms to predict the stability of perovskite oxides. Our machine learning models were trained on a DFT-calculated dataset consisting of 1929 compounds, and proved to be a promising tool to successfully predict $E_{hull}$ values within typical DFT calculation errors. We constructed a set of 791 features by considering combinations of the properties of the elements comprising each perovskite material, and used feature selection routines to select the top 70 features result in the best predictions of material stability without significant overfitting. We selected the extra trees classifier as the best model for classification and the kernel ridge regression as the best model for regression from five candidate models, by comparing the performance in cross-validation. The best $F_1$ score achieved for classification was 0.881 (+/- 0.032) and the best RMSE value for regression of $E_{hull}$ was 28.5 (+/- 7.5) meV/atom. We validated our model by performing select extrapolation tests on five subsets using a targeted cross-validation scheme, and our model showed good classification and regression performance in these test cases, even when the composition information in the training dataset is limited. Furthermore, we applied our models on new, manually generated perovskite compounds and compared them with DFT calculations. The model was able to give close predictions for compounds containing elements frequently sampled in the training dataset, which makes this model useful in predicting the stability of perovskite oxides in the composition space relevant for high activity SOFC cathodes. Considering the fast prediction speed, our models show potential for fast screening of new candidate materials in a large composition space via machine learning approaches.

**Conflicts of interest**

There are no conflicts to declare.

**Data Availability**

The dataset of the thermodynamic stability of 1929 perovskite oxides, elemental properties of all element in periodic table, as well as the machine learning model package for predicting stability of new oxides are provided in **Data in Brief (DiB)**.

**Acknowledgements**

Funding for this work was provided by the NSF Software Infrastructure for Sustained Innovation (SI2) award No. 1148011. Computational support was provided by the Extreme Science and Engineering Discovery Environment (XSEDE), which is supported by National Science Foundation Grant No. OCI-1053575.

Reference

1. Ishihara, T., *Perovskite oxide for solid oxide fuel cells*. 2009: Springer Science & Business Media.
2. Jacobs, R., et al., *Material Discovery and Design Principles for Stable, High Activity Perovskite Cathodes for Solid Oxide Fuel Cells.* Advanced Energy Materials, 2018: p. 1702708.
3. Emery, A., et al., *High-Throughput Computational Screening of Perovskites for Thermochemical Water Splitting Applications.* Chemistry of Materials, 2016. **28**(16).
4. Wang, S., et al., *Assessing the thermoelectric properties of sintered compounds via high-throughput ab-initio calculations.* Physical Review X, 2011. **1**(2): p. 021012.
5. Carrete, J., et al., *Finding unprecedentedly low-thermal-conductivity half-Heusler semiconductors via high-throughput materials modeling.* Physical Review X, 2014. **4**(1): p. 011019.
6. Castelli, I.E., et al., *Computational screening of perovskite metal oxides for optimal solar light capture.* Energy & Environmental Science, 2012. **5**(2): p. 5814-5819.
7. Wu, Y., et al., *First principles high throughput screening of oxynitrides for water-splitting photocatalysts.* Energ. Environ. Sci., 2013. **6**(1): p. 157-168.
8. Castelli, I.E., et al., *New cubic perovskites for one-and two-photon water splitting using the computational materials repository.* Energy & Environmental Science, 2012. **5**(10): p. 9034-9043.
9. Greeley, J., et al., *Computational high-throughput screening of electrocatalytic materials for hydrogen evolution.* Nat Mater, 2006. **5**(11): p. 909-913.
10. GreeleyJ, et al., *Alloys of platinum and early transition metals as oxygen reduction electrocatalysts.* Nat Chem, 2009. **1**(7): p. 552-556.

11. Scott, D., et al., *Prediction of the functional properties of ceramic materials from composition using artificial neural networks.* Journal of the European Ceramic Society, 2007. **27**(16): p. 4425-4435.
12. Lee, J., et al., *Prediction model of band gap for inorganic compounds by combination of density functional theory calculations and machine learning techniques.* Physical Review B, 2016. **93**(11): p. 115104.
13. Faber, F.A., et al., *Machine Learning Energies of 2 Million Elpasolite (A B C 2 D 6) Crystals.* Physical review letters, 2016. **117**(13): p. 135502.
14. Montavon, G., et al., *Machine learning of molecular electronic properties in chemical compound space.* New Journal of Physics, 2013. **15**(9): p. 095003.
15. Schütt, K., et al., *How to represent crystal structures for machine learning: Towards fast prediction of electronic properties.* Physical Review B, 2014. **89**(20): p. 205118.
16. Hansen, K., et al., *Assessment and validation of machine learning methods for predicting molecular atomization energies.* Journal of Chemical Theory and Computation, 2013. **9**(8): p. 3404-3419.
17. Armiento, R., et al., *High-throughput screening of perovskite alloys for piezoelectric performance and thermodynamic stability.* Physical Review B, 2014. **89**(13): p. 134103.
18. Meredig, B., et al., *Combinatorial screening for new materials in unconstrained composition space with machine learning.* Physical Review B, 2014. **89**(9): p. 094104.
19. Ward, L., et al., *A general-purpose machine learning framework for predicting properties of inorganic materials.* Npj Computational Materials, 2016. **2**: p. 16028.
20. Schmidt, J., et al., *Predicting the thermodynamic stability of solids combining density functional theory and machine learning.* Chemistry of Materials, 2017.
21. Liu, M., et al., *Spinel compounds as multivalent battery cathodes: a systematic evaluation based on ab initio calculations.* Energy & Environmental Science, 2015. **8**(3): p. 964-974.
22. Jain, A., et al., *Commentary: The Materials Project: A materials genome approach to accelerating materials innovation.* Apl Materials, 2013. **1**(1): p. 011002.
23. Geurts, P., D. Ernst, and L. Wehenkel, *Extremely randomized trees.* Machine learning, 2006. **63**(1): p. 3-42.
24. Hoerl, A.E. and R.W. Kennard, *Ridge regression: Biased estimation for nonorthogonal problems.* Technometrics, 1970. **12**(1): p. 55-67.
25. Pedregosa, F., et al., *Scikit-learn: Machine learning in Python.* Journal of Machine Learning Research, 2011. **12**(Oct): p. 2825-2830.
26. Li, W., R. Jacobs, and D. Morgan, *Data and Supplemental information for Predicting the thermodynamic stability of perovskite oxides using machine learning models.* Data in Brief, 2018. **19**: p. 261-263.
27. Sun, W., et al., *The thermodynamic scale of inorganic crystalline metastability.* Science Advances, 2016. **2**(11).
28. Herr, N. *The Sourcebook for Teaching Science*. 2007 [cited accessed May 2017; Available from: https://www.csun.edu/science/ref/spreadsheets/.
29. Goldschmidt, V.M., *Die gesetze der krystallochemie.* Naturwissenschaften, 1926. **14**(21): p. 477-485.
30. Li, C., K.C.K. Soh, and P. Wu, *Formability of ABO 3 perovskites.* Journal of alloys and compounds, 2004. **372**(1): p. 40-48.

31. Shannon, R.t., *Revised effective ionic radii and systematic studies of interatomic distances in halides and chalcogenides.* Acta crystallographica section A: crystal physics, diffraction, theoretical and general crystallography, 1976. **32**(5): p. 751-767.
32. Meinshausen, N. and P. Bühlmann, *Stability selection.* Journal of the Royal Statistical Society: Series B (Statistical Methodology), 2010. **72**(4): p. 417-473.
33. Guyon, I., et al., *Gene selection for cancer classification using support vector machines.* Machine learning, 2002. **46**(1): p. 389-422.
34. Kraskov, A., H. Stögbauer, and P. Grassberger, *Estimating mutual information.* Physical review E, 2004. **69**(6): p. 066138.
35. Hosmer Jr, D.W., S. Lemeshow, and R.X. Sturdivant, *Applied logistic regression*. Vol. 398. 2013: John Wiley & Sons.
36. Yegnanarayana, B., *Artificial neural networks*. 2009: PHI Learning Pvt. Ltd.
37. Suykens, J.A. and J. Vandewalle, *Least squares support vector machine classifiers.* Neural processing letters, 1999. **9**(3): p. 293-300.
38. *Model evaluation: quantifying the quality of predictions*. [cited accessed May 2017; Available from: http://scikit-learn.org/stable/modules/model_evaluation.html.
39. Saal, J.E., et al., *Materials design and discovery with high-throughput density functional theory: the open quantum materials database (OQMD).* Jom, 2013. **65**(11): p. 1501-1509.
40. Gaultois, M.W., et al., *Data-driven review of thermoelectric materials: performance and resource considerations.* Chemistry of Materials, 2013. **25**(15): p. 2911-2920.
41. Park, S.H., J.M. Goo, and C.-H. Jo, *Receiver operating characteristic (ROC) curve: practical review for radiologists.* Korean Journal of Radiology, 2004. **5**(1): p. 11-18.
42. Jacobson, A.J., *Materials for solid oxide fuel cells.* Chemistry of Materials, 2009. **22**(3): p. 660-674.
43. Haile, S.M., *Fuel cell materials and components.* Acta Materialia, 2003. **51**(19): p. 5981-6000.